\documentclass[twocolumn,showpacs,preprintnumbers,nofootinbib,aps,prd,superscriptaddress,longbibliography,10pt]{revtex4-1}

\usepackage{graphicx,amssymb,amsmath,amsthm,amsfonts,epsfig,times}
\usepackage[linktocpage]{hyperref}
\usepackage[usenames,dvipsnames]{color}
\usepackage{epstopdf}
\usepackage{enumerate}
\usepackage{cleveref}
\definecolor{coolblack}{rgb}{0.0, 0.18, 0.39}
\definecolor{darkred}{rgb}{0.5,0,0}
\definecolor{darkgreen}{rgb}{0,0.5,0}
\definecolor{darkblue}{rgb}{0,0,0.5}
\definecolor{lapislazuli}{rgb}{0.15, 0.38, 0.61}
\definecolor{venetianred}{rgb}{0.78, 0.03, 0.08}
\definecolor{bleudefrance}{rgb}{0.19, 0.55, 0.91}
\definecolor{dogwoodrose}{rgb}{0.84, 0.09, 0.41}
\definecolor{dogwoodrose}{rgb}{0.84, 0.09, 0.41}
\hypersetup{colorlinks=true, citecolor=darkblue, linkcolor=darkblue, 
urlcolor = darkblue}

\usepackage{amsmath,amssymb}
\usepackage{tensor}

\def\be{\begin{equation}}
\def\ee{\end{equation}}

\def\ss{\scriptscriptstyle}

\newcommand{\bea}{\begin{eqnarray}}
\newcommand{\eea}{\end{eqnarray}}
\newcommand{\ben}{\begin{enumerate}}
\newcommand{\een}{\end{enumerate}}
\newcommand{\bi}{\begin{itemize}}
\newcommand{\ei}{\end{itemize}}

\def\ga{\mathrel{\raise.3ex\hbox{$>$\kern-.75em\lower1ex\hbox{$\sim$}}}}
\def\la{\mathrel{\raise.3ex\hbox{$<$\kern-.75em\lower1ex\hbox{$\sim$}}}}

\def\l{\left}
\def\r{\right}
\def\be{\begin{equation}}
\def\ee{\end{equation}}

\def\ss{\scriptscriptstyle}

\def\I_M{{I_{\scriptscriptstyle M\times M}}}

\def\be{\begin{equation}}
\def\ee{\end{equation}}
\def\bea{\begin{eqnarray}}
\def\eea{\end{eqnarray}}
\newcommand{\beq}{\begin{eqnarray}}
\newcommand{\eeq}{\end{eqnarray}}

\begin{document}\title{\large Quasinormal modes of relativistic stars and interacting fields}

\author{Caio F. B. Macedo}\email{caio.macedo@tecnico.ulisboa.pt}
\affiliation{CENTRA, Departamento de F\'{\i}sica, Instituto Superior T\'ecnico--IST, Universidade de Lisboa--UL, Avenida~Rovisco Pais 1, 1049 Lisboa, Portugal}
\affiliation{Faculdade de F\'{\i}sica, Universidade 
Federal do Par\'a, 66075-110 Bel\'em, Par\'a, Brazil}

\author{Vitor Cardoso}\email{vitor.cardoso@ist.utl.pt}
\affiliation{CENTRA, Departamento de F\'{\i}sica, Instituto Superior T\'ecnico--IST, Universidade de Lisboa--UL, Avenida~Rovisco Pais 1, 1049 Lisboa, Portugal}
\affiliation{Perimeter Institute for Theoretical Physics, Waterloo, Ontario N2J 2W9, Canada}

\author{Lu\'is C. B. Crispino}\email{crispino@ufpa.br}
\affiliation{Faculdade de F\'{\i}sica, Universidade 
Federal do Par\'a, 66075-110 Bel\'em, Par\'a, Brazil}

\author{Paolo Pani}\email{paolo.pani@roma1.infn.it}
\affiliation{Dipartimento di Fisica, ``Sapienza'' Universit\`a di Roma and Sezione INFN Roma1, Piazzale Aldo Moro 5, 00185 Roma, Italy}
\affiliation{CENTRA, Departamento de F\'{\i}sica, Instituto Superior T\'ecnico--IST, Universidade de Lisboa--UL, Avenida~Rovisco Pais 1, 1049 Lisboa, Portugal}
\begin{abstract}
The quasinormal modes of relativistic compact objects encode important information about the gravitational response associated with astrophysical phenomena. Detecting such oscillations would provide us with a unique understanding of the properties of compact stars, and may give definitive evidence for the existence of black holes. However, computing quasinormal modes in realistic astrophysical environments is challenging due to the complexity of the spacetime background and of the dynamics of the perturbations.
We discuss two complementary methods for computing the quasinormal modes of spherically symmetric astrophysical systems, namely, the direct integration method and the continued-fraction method. {We extend these techniques to dealing with generic coupled systems of linear equations, with the only assumption being} the interaction between different fields is effectively localized within a finite region. {In particular, we adapt the continued-fraction method to include cases where a series solution can be obtained only outside an effective region. As an application}, we compute the polar quasinormal modes of boson stars {by using the continued-fraction method for the first time.} The methods discussed here can be applied to other situations in which the perturbations effectively couple only within a finite region of space.

\end{abstract}


%
%

\maketitle


\section{Introduction}
\label{sec:intro}

The study of the natural oscillations of physical systems is of much interest in astrophysics~\cite{Berti:2009kk,Kokkotas:1999bd}. For relativistic compact objects such as stars and black holes, these natural oscillations are referred to as quasinormal modes (QNMs). QNMs depend only on the properties of the object. The radiation emitted in the collapse of a star and in the coalescence of binaries has a close relation to the QNMs of the final compact object. The composition of matter forming astrophysical objects is very important for the dynamics of the spacetime oscillations. Indeed, many matter oscillations in Newtonian theory are also necessarily present in the relativistic case~\cite{Kokkotas:1999bd}. Fluid modes can also be excited, for instance, by a particle moving around the compact object~\cite{Kojima:1987tk,Pons:2001xs}, generating a large dephasing in the gravitational-wave signal~\cite{Macedo:2013qea}.

It is also of great interest to study how matter and fields around black holes behave. The presence of matter around black holes can lead to resonances, which were analyzed in detail in Ref.~\cite{Barausse:2014tra}. Scalar fields around rotating Kerr black holes can generate hairy configurations~\cite{Herdeiro:2014goa} (see also Refs.~\cite{Herdeiro:2015waa,Brito:2015oca} for a recent review of black holes with scalar hair). Moreover, massive scalar fields can form quasibound states and develop scalar ``clouds'' around black holes~\cite{Hod:2012px,Benone:2014ssa,Barranco:2012qs,Zilhao:2015tya,Brito:2015oca}. Notwithstanding, relativistic stars can also present nontrivial gravitating field configurations. For instance, very strong magnetic fields in neutron stars---configurations known as magnetars~\cite{Duncan:1992hi}---have an influence on the shape of the star~\cite{Konno:1999zv,Yazadjiev:2011ks}. Neutron stars can also accrete dark matter, acquiring a core formed by an additional gravitating component~\cite{Bertone:2007ae,Brito:2015yga,Brito:2015yfh}. Moreover, stars formed by fundamental fields, like boson stars (BSs)~\cite{Kaup:1968zz,Schunck:2003kk,Liebling:2012fv} and  Proca stars~\cite{Brito:2015pxa}, are also interesting examples of the outcomes of self-interacting, gravitating fields.

All of the above examples share a common feature: they are formed by components (matter and fields) coupled to each other, with such a coupling being crucial for the structure of the compact object. These interactions usually have a characteristic length scale so that, at large distances, all fields decouple or only the gravitational field survives (through a power-law falloff). 

The coupling among different fields also impacts the analysis of linear perturbation in such spacetimes. Additional components describing the perturbations of relativistic objects can enrich their oscillation spectra, generating distinctive signatures. For instance, the scalar field modes in BSs can be excited by an orbiting particle~\cite{Macedo:2013qea}. Notwithstanding, the scattering of massive scalar wave packets can carry information about the clouds around black holes~\cite{Degollado:2014vsa}. However, it is difficult to analyze the QNM spectrum of such systems, basically due to the fact that the coupling between the components is highly nontrivial in most cases. 

Here, we {present} two distinct methods for computing the QNMs of relativistic systems---namely, the direct integration method and the continued fraction method---to deal with situations in which the linearized perturbations interact only in a finite region of space. The direct integration method discussed here is {similar to} the one presented in Ref.~\cite{Pani:2013pma}. On the other hand, we extend the continued-fraction method adopted, e.g., in Refs.~\cite{Benhar:1998au,Nollert:1999ji}, to deal with arbitrary coupled systems of linear equations. 
{One of the main advantages of our improved method is that it does not need to assume a continued-fraction representation that describes the solutions in the entire domain. Instead, we only require that a continued-fraction representation exists outside an effective interaction region. This is an advantage because, in many physical situations, the metric is analytically known only outside an effective radius, so that a continued-fraction representation is available only in a portion of the full domain.}
We shall assume that the angular part of the perturbations can be separated such that we end up with a system of $N$ coupled second order differential equations.
This separation naturally occurs in spherically symmetric configurations, and also for spinning, axisymmetric objects in the slow-rotation approximation~\cite{Pani:2013pma}.
Additionally, we assume that the background is stationary, so that the perturbations can be Fourier decomposed with time dependence $e^{-i\sigma t}$, where $\sigma$ is the Fourier frequency. The above assumptions are valid in a wide variety of astrophysical scenarios~\cite{Berti:2009kk,Pani:2013pma,Barausse:2014tra}.

{The decoupling of the perturbations outside an effective radius occurs naturally in BSs. In fact, the coupling between the gravitational and scalar field perturbations decrease exponentially with the radial coordinate and become negligible beyond a certain radius, where the scalar field which composes the BSs becomes small. Therefore, as an application of our improved methods, we compute the QNMs of mini-BSs through the continued-fraction technique, extending previous results obtained with less robust methods~\cite{Macedo:2013jja}. We compute the polar QNMs of mini-BSs as functions of the central field density and show that monopolar modes become unstable beyond a critical density which corresponds to the maximum mass of the BS.}

The remainder of this paper is organized as follows: in Sec. \ref{sec:methods} we describe the direct integration and continued-fraction methods for computing the QNMs modes of compact objects surrounded by interacting fields. In Sec. \ref{sec:qnmboson} we apply the improved direct integration and continued-fraction methods to compute the QNMs of BSs. Moreover, we extend the results presented in Ref. \cite{Macedo:2013jja} by computing the QNMs of mini BSs as a function of the central field and compactness. We show that the $l=0$ modes become unstable beyond a critical density corresponding to the maximum mass of the star. In Sec.~\ref{sec:discusscf} we discuss our results and deliver our final remarks.

\section{Methods}\label{sec:methods}

We assume that the perturbation functions can be described by the following set of equations:
\bea
\frac{d^2}{dr_*^2}\mathbf{\Psi}(r)+(\sigma^2-\mathbf{V}(r))\mathbf{\Psi}(r)=0,
\label{eq:CFeqs}
\eea
where $r_*$ is a tortoise radial coordinate, $\sigma$ is the frequency of the field, $\mathbf{\Psi}(r)$ is an $N$-dimensional vector representing the perturbations, and $\mathbf{V}(r)$ is an $N\times N$ matrix which can possibly depend on $\sigma$. If the components of the perturbation $\mathbf{\Psi}(r)$ are decoupled, $\mathbf{V}(r)$ is a diagonal matrix. In some scenarios, it may be possible to obtain a set of decoupled equations for the perturbations by performing a canonical transformation within a Hamiltonian framework~\cite{Brizuela:2015zia}. However, the only restriction to ${\bf V}(r)$ here is that it has a diagonal form for $r> l_i$, where $l_i$ denotes the characteristic range of the interaction between the components of ${\bf \Psi}(r)$.

\subsection{Direct integration method}
 This method consists of integrating the differential equations within two different regions: near the origin and far from the star, with the proper QNM boundary conditions. The problem of finding the QNM frequencies reduces to finding the proper values of $\sigma$ for which the solutions obtained integrating from the origin and from infinity are linearly dependent. This method was first used by Chandrasekhar and Detweiler~\cite{Chandrasekhar:1975zza}, and it was then applied to uniform density stars to compute the least damped (lowest imaginary part) QNMs~\cite{chandrasekharferrariIII}.
 
The boundary conditions at the origin can be written as\footnote{The potential ${\bf V} (r)$ usually diverges at the origin due to a centrifugal term, and that is the reason for the $r^{l}$ term in Eq.~\eqref{eq:expansionor}, where $l$ is the angular number of the waves. See, e.g., Ref.~\cite{Cardoso:2014sna} and the references therein. We note that, if the central object is a black hole, QNMs are defined by ingoing waves at the event horizon. In this case, a different set of boundary conditions should be imposed in place of Eq.~\eqref{eq:expansionor}.}
\be
\mathbf{\Psi}(r\sim 0)\sim r^l \sum_{i=0}^{N_0} \mathbf{x}_0^i~ r^{i}\,,
\label{eq:expansionor}
\ee
where the	 $\mathbf{x}_0^i$ are constant $N$-dimensional vectors and the upper summation limit $N_0$ is chosen such that the boundary conditions converge to the required accuracy. Substituting Eq.~\eqref{eq:expansionor} into Eq.~\eqref{eq:CFeqs} and expanding around $r\sim 0$ leads to a recursion relation for the coefficients $\mathbf{x}_0^i$, such that all of them can be written as functions of the coefficients $\mathbf{x}_0^0$, which is a collection of $N$ independent numbers. In this way, since the system is linear, we can form a set of $N$ independent solutions by integrating from the origin by choosing the vector $\mathbf{x}^0_0$ to be, e.g., $(1,0,...,0),~(0,1,...,0),...$, and $(0,0,...,1)$.	The general solution can be achieved by a linear combination of the $N$ independent solutions, namely,
\be
\mathbf{\Psi}^{-}(r)=\sum_{n=1}^{N}\alpha_n^- ~\mathbf{\Psi}^-_n(r),
\label{eq:indor}
\ee
where $\mathbf{\Psi}_{n}^-$ denotes the $n$th independent solution of Eq.~\eqref{eq:CFeqs}, obtained by integrating it from the origin. 

The boundary conditions at infinity can be written as
\be
{\Psi_j}(r\sim \infty)\sim \exp({\pm { k_j}(\sigma)\, r_*})\sum_{i=0}^{N_0} \frac{{x}_{j,\infty}^{i} }{r^{i}}	\,,
\label{eq:expansioninf}
\ee
where 
\be
k_j(\sigma)=\sqrt{V_{jj}(r\to\infty)-\sigma^2}.
\ee
Depending on $V_{jj}(r\to\infty)$ and on the value of the frequency $\sigma$,  we select the value of the $\pm$ sign in Eq.~\eqref{eq:expansioninf} to suit the particular problem. For example, when $\sigma^2>V_{jj}(r\to\infty)$, the system allows perturbations which are wavelike at infinity, $\Psi_j\sim e^{\pm k_j(\sigma)\, r_*}$ {[with $k_j(\sigma)$ being complex]}; when $\sigma^2<V_{jj}(r\to\infty)$, the perturbations are bounded and we require an exponential damping $\Psi_j\sim e^{-k_j(\sigma)\, r_*}$ (see, e.g., Table~II of Ref.~\cite{Macedo:2013jja}).  The $\mathbf{x}_{\infty}^i$ are constant $N$-dimensional vectors and the upper summation limit $N_0$ in Eq.~\eqref{eq:expansioninf} has to be chosen according to the precision required. 
Similar to the integration from the origin discussed above, one can construct a set of $N$ independent solutions by integrating the perturbation equations from infinity and by choosing different values for $\mathbf{x}_{\pm,\infty}^0$. The solution integrated from infinity can be written as the following linear combination:
\be
\mathbf{\Psi}^{+}(r)=\sum_{n=0}^{N}\alpha_n^+ ~\mathbf{\Psi}^+_n(r),
\label{eq:psiout}
\ee
where $\mathbf{\Psi}_{n}^+$ denotes the $n$th independent solution of Eq.~\eqref{eq:CFeqs}.
The QNM solutions are such that $\mathbf{\Psi}^{-}(r)$ and $\mathbf{\Psi}^{+}(r)$ are linearly dependent. We see that the QNM frequencies can be found through
\be
\begin{array}{c}
\l.\mathbf{\Psi}^{-}(r)\r|_{r=R_{\rm ext}}=\l.\mathbf{\Psi}^{+}(r)\r|_{r=R_{\rm ext}},\\
\l.{\mathbf{\Psi}^{-}}'(r)\r|_{r=R_{\rm ext}}=\l.{\mathbf{\Psi}^{+}}'(r)\r|_{r=R_{\rm ext}}	,
\end{array}
\label{eq:continuityi}
\ee
where $R_{\rm ext}>l_i$. The conditions~\eqref{eq:continuityi} generate a system of $2N$ equations for the $2N$ coefficients $\alpha_n^\pm$ and for $\sigma$. Since the system is linear, we can set one of the $\alpha_n^\pm$'s to unity; say, for instance, $\alpha_0^+ =1$. We then use the remaining $2N-1$ equations to find the rest of the coefficients as functions of $\sigma$. The remaining equation is then used to find the QNM frequencies.

\subsection{Continued-fraction method}\label{sec:DI}
The continued-fraction method is a very powerful technique, with many applications in physics. In the context of QNMs, it was first used by Leaver~\cite{Leaver:1985ax}, and has been extensively studied by many authors since then~\cite{Berti:2009kk}. One of the main difficulties in computing QNMs in the frequency domain is the divergence of the wave functions at large distances. The continued-fraction method works extremely well in some cases because it maps the divergent boundary behavior in a specific recurrence relation.

For stellar structures, the continued-fraction method should actually be used in combination with a direct integration. Outside the star (in the vacuum region), one can recast the solution in terms of a continued fraction. 
However, in order to guarantee that all the proper boundary conditions are satisfied, we have to match continuously this outer solution with the one obtained by integrating the differential equations from the origin directly.

{Reference~\cite{Pani:2013pma} presents a method for computing the modes through continued fractions, assuming that one can find a series representation in the entire domain. In particular, the method of Ref.~\cite{Pani:2013pma} requires that the background spacetime is known analytically. However, in many astrophysical scenarios the metric is constructed only numerically, which makes it difficult to solve the equations through a Frobenius method in the entire domain. Fortunately, even in the presence of interacting fluids and fields, the spacetime might still be described analytically, at least outside an effective radius. This is the case of (slowly rotating) stellar configurations, black holes surrounded by matter, or self-gravitating solitons like BSs. In these cases, it is possible to obtain a continued-fraction representation of the solutions outside an effective radius, and then to match this solution with the one obtained by integrating the equations from requiring regularity at the origin. In this way, we can assure that both boundary conditions are satisfied.}

{Let us expose the procedure to obtain a series representation outside the effective radius.} We again assume that the system of equations can be described by Eq.~\eqref{eq:CFeqs} and that there exists an expansion of the components of the wave vector as follows:
\be
\Psi_i^{+}(r)=\Xi_i(r)\sum_{n=0}^\infty a_{i,n}~ {v}^n,
\label{eq:CFexpansion}
\ee
where the index $i$ denotes the $i$th component of the vector functions and  $v\equiv (1-b/r)$, in which $b$ is chosen such that the series solution \eqref{eq:CFexpansion} is convergent~\cite{Benhar:1998au}.
The vector function $\mathbf{\Xi}$ is chosen such that the vector $\mathbf{\Psi}^+$ satisfies the proper boundary conditions at infinity~\cite{RevModPhys.83.793}---typically, its components are proportional to $e^{\pm k_j(\sigma)r_*}$. Substituting the expansion~\eqref{eq:CFexpansion} into the differential equation~\eqref{eq:CFeqs} leads to recurrence relations for the coefficients $a_{i,n}$~\cite{Leaver:1985ax,Pani:2013pma}. These could, in principle, be $n$-term recurrence relations, which can be reduced to a three-term recurrence relation by recursive Gaussian elimination steps. {The recurrence can be solved, writing all the coefficients $a_{i,n}$---with $n>0$---in terms of $a_{i,0}$. Bellow, we explain the procedure.}

Let us illustrate this by assuming that we end up with a four-term recurrence relation. We have
\be
\boldsymbol{\alpha}_n \mathbf{a}_{n+1}+\boldsymbol{\beta}_n\mathbf{a}_n+\boldsymbol{\gamma}_{n}\mathbf{a}_{n-1}+\boldsymbol{\delta}_n\mathbf{a}_{n-2}=0,~~~~n>1,
\label{eq:fourterm}
\ee
where $\boldsymbol{\alpha}_n$, $\boldsymbol{\beta}_n$, $\boldsymbol{\gamma}_n$ and $\boldsymbol{\delta}_n$ are $N\times N$ invertible matrices. $\mathbf{a}_{n}$ is the vector whose components are the coefficients $a_{i,n}$ in the expansion~\eqref{eq:CFexpansion}. We can reduce the four-term recurrence relation to a three-term one by using a matrix-valued Gaussian elimination step~\cite{Pani:2013pma}. Using
\bea
\tilde{\boldsymbol{\alpha}}_n&=&{\boldsymbol{\alpha}}_n,\\
\tilde{\boldsymbol{\beta}}_0&=&{\boldsymbol{\beta}}_0,\\
\tilde{\boldsymbol{\gamma}}_0&=&{\boldsymbol{\gamma}}_0,\\
\tilde{\boldsymbol{\beta}}_n&=&{\boldsymbol{\beta}}_n-{\boldsymbol{\delta}}_n\l[\tilde{\boldsymbol{\gamma}}_{n-1}\tilde{\boldsymbol{\alpha}}_{n-1}\r]^{-1},~~~n>0,\\
\tilde{\boldsymbol{\gamma}}_n&=&{\boldsymbol{\gamma}}_n-{\boldsymbol{\delta}}_n\l[\tilde{\boldsymbol{\gamma}}_{n-1}\tilde{\boldsymbol{\alpha}}_{n-1}\r]^{-1},~~~n>0,
\eea
one can show, through Eq.~\eqref{eq:fourterm}, that the tilde matrices satisfy the following three-term recurrence relation:
\be
\tilde{\boldsymbol{\alpha}}_n \mathbf{a}_{n+1}+\tilde{\boldsymbol{\beta}}_n\mathbf{a}_n+\tilde{\boldsymbol{\gamma}}_{n}\mathbf{a}_{n-1}=0,~~~~n>0.
\label{eq:threeterm}
\ee
Defining a ladder matrix $\mathbf{R}_n^+$ with the following property,
\be
\mathbf{a}_{n+1}=\mathbf{R}_n^+ \mathbf{a}_{n},
\label{eq:aladder}
\ee
and using Eq.~\eqref{eq:threeterm}, we obtain the following equation:
\be
\mathbf{R}_n^+=-\l[\tilde{\boldsymbol{\beta}}_{n+1}+\tilde{\boldsymbol{\alpha}}_{n+1}\mathbf{R}_{n+1}^+\r]^{-1}\boldsymbol{\gamma}_{n+1}.
\label{eq:ladderR}
\ee
Equation~\eqref{eq:ladderR} may be solved recursively. We can start at some large value of $n$---say $N_0$---impose $\mathbf{R}_{N_0}^+=0$, and then, moving backwards in $n$, determine all $\mathbf{R}_n^+$'s. The result has the form of a continued fraction, justifying the name of the method.\footnote{For instance, if we have just one second order equation, the solution of the recurrence relation~\eqref{eq:ladderR} takes the form
\be
R^{+}_n=-\frac{\tilde{\gamma}_{n+1}}{\tilde{\beta}_{n+1}-}\frac{\tilde{\alpha}_{n+1}\tilde{\gamma}_{n+2}}{\tilde{\beta}_{n+2}-}\frac{\tilde{\alpha}_{n+2}\tilde{\gamma}_{n+3}}{\tilde{\beta}_{n+3}-}\cdots,
\ee
where we used the notation $\frac{a}{b-}\frac{c}{d}\equiv\frac{a}{b-\frac{c}{d}}$.
}

Using Eqs.~\eqref{eq:aladder} and~\eqref{eq:ladderR}, we obtain all $\mathbf{a}_{n}$'s as functions of $N$ parameters, given by $\mathbf{a}_{0}$. The expansion \eqref{eq:CFexpansion}, with the determined coefficients, gives the solution to be used outside the star. Therefore, there are $N$ independent solutions of Eq.~\eqref{eq:CFeqs}, and the general solution is a linear combination of them, similar to Eq.~\eqref{eq:psiout}. 
{Note that, since the recurrence relations are valid outside the effective radius, the solutions in the interior cannot be described by this procedure.}

We still have to impose that the wave function $\mathbf{\Psi}$ satisfies the proper boundary condition at the origin. For this purpose, we can use the same procedure as in Sec.~\ref{sec:DI}, obtaining ${\bf \Psi}^{-}(r)$, given by Eq.~\eqref{eq:indor}.

The QNM frequencies are found by requiring that the wave functions obtained from inside and outside be linearly dependent.
Similarly to the direct integration method, we impose Eq.~\eqref{eq:continuityi}. Note that the difference between the two methods is that we construct the outer solutions ${\bf \Psi}^{+}$ through the continued-fraction approximations. For relativistic ordinary stars, the continued-fraction method presented above reduced to the one presented in Refs.~\cite{Benhar:1998au,Nollert:1999ji}.


\section{Application: QNMs of a Boson star}\label{sec:qnmboson}

\begin{table*}
\caption{Comparison of $l=2$ QNMs of a BSs computed through the direct integration (DI) and continued-fraction (CF) methods. In this particular case, the direct integration method is the same as the one used in Ref.~\cite{Macedo:2013qea}. \label{tab:modes}}
\begin{tabular}{l l l | l l} 
\hline \hline
Configurations of Ref.~\cite{Macedo:2013jja} & DI & & CF & \\
\hline
Mini-BS I, N=1 & $0.1195$ & $-5\times10^{-5} i$ & $0.1186$ & $-5.3\times10^{-5}i$ \\
Mini-BS I, N=2 & $0.1316$ & $-2\times10^{-5} i$ & $0.1316$ & $-2.4\times10^{-6}i$ \\
Mini-BS I, N=3 & $0.1404$ & $-8\times 10^{-6} i$ & $0.1404$ & $-7.7\times 10^{-6} i$\\
Massive-BS I, N=1 & $4.03\times 10^{-2}$ & $-2\times 10^{-5} i$ & $4.029\times 10^{-2}$ & $-2.5\times 10^{-5} i$ \\
Massive-BS I, N=2 & $7.16\times 10^{-2}$ & $-2\times 10^{-6} i$ & $7.158\times 10^{-2}$ & $-2.1\times 10^{-6} i$ \\
Massive-BS I, N=3 & $9.47\times 10^{-2}$ & $-5\times 10^{-7} i$ & $9.465\times 10^{-2}$ & $-4.7	\times 10^{-7} i$ \\
\hline\hline
\end{tabular}
\end{table*}

Here we extend the calculations of Ref.~\cite{Macedo:2013jja}, computing the QNMs of a BS through the improved continued-fraction method. BSs are starlike configurations formed by self-interacting gravitating complex scalar fields~\cite{Kaup:1968zz,Schunck:2003kk}. In the case of BSs, outside an effective radius, the perturbations of the spacetime are well described by a decoupled system---two equations describing the polar and axial parts of gravitational perturbations and two equations describing the perturbations of the complex scalar field.\footnote{As discussed in Ref.~\cite{Macedo:2013jja}, although the two equations governing the perturbations for the scalar field are coupled to each other even outside the star, one can always decouple them by a suitable field redefinition~\cite{Gleiser:1988rq}.} The external spacetime is very well approximated by the Schwarzschild geometry and, therefore, the continued-fraction expressions for the wave functions outside the star are the same as in the Schwarzschild case. For the gravitational part of the perturbations, one can construct the continued-fraction relation using the Regge-Wheeler equation and use it to obtain the polar and axial gravitational perturbations~\cite{Leaver:1985ax,Chandrasekhar:1985kt}. For the scalar field perturbations, we can use the results of Ref.~\cite{Konoplya:2004wg}.

For the axial sector of the perturbations, the application of the above method is direct since the equation for the axial perturbations is already in the form of Eq.~\eqref{eq:CFeqs}. Moreover, the scalar field perturbations only couple to the polar sector of the perturbations~\cite{Kojima:1991np,Yoshida:1994xi}.

For the polar sector of the perturbations, the method can be applied using the first order differential equations for the gravitational perturbations presented in Ref.~\cite{Macedo:2013jja} in the following way.
\begin{enumerate}[i]
	\item We solve the differential equations from the origin, choosing the three free parameters such that we obtain the three independent solutions for the metric and scalar field perturbations.
	\item Using the perturbations obtained by the procedure above, we construct the corresponding Zerilli function $\Psi_Z$ (which describes the polar gravitational perturbations) of each independent solution. With the independent solutions $(\Psi_Z,\phi_+,\phi_-)$, we obtain ${\bf \Psi}^{-}(r)$ [Eq.~\eqref{eq:indor}] (see Ref.~\cite{Macedo:2013jja} for more details).
	\item The vector $\mathbf{\Psi}^{+}(r)$ can be constructed using either the continued-fraction or the direct integration method. Since the gravitational and scalar parts of the perturbations are independent, they can be computed separately;
	\item Finally, with the vector $\mathbf{\Psi}^{+}(r)$, we find the QNMs by matching it at $r=R_{\rm ext}$ with ${\bf \Psi}^{-}(r)$, through Eq.~\eqref{eq:continuityi}.
\end{enumerate}
Note that, for the polar case, since we are using the Zerilli function to describe the gravitational perturbations, the point $R_{\rm ext}$ should be such that the background scalar field at that point is negligible.

\begin{figure}
\hspace*{-1cm}
\includegraphics[width=0.65\textwidth]{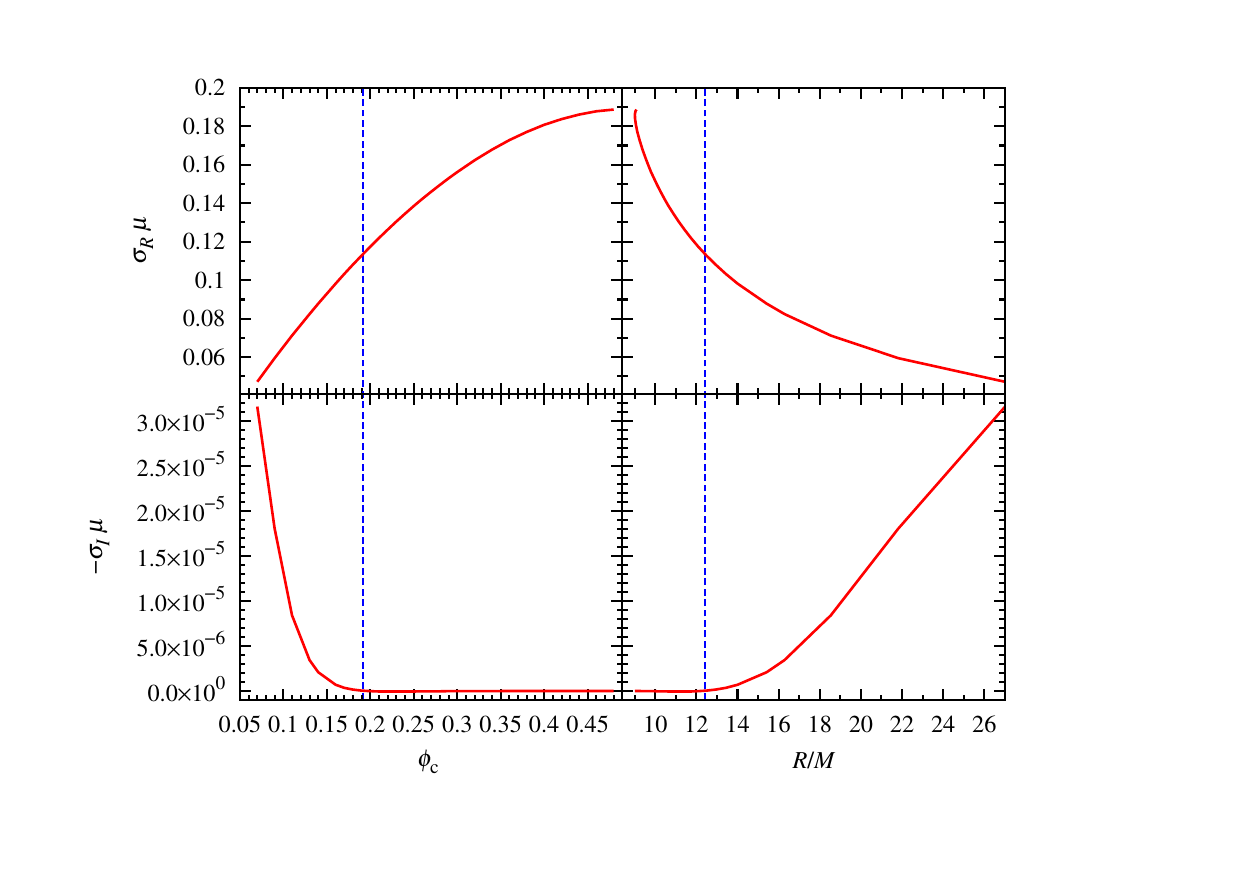}%
\vspace*{-1cm}
\caption{Real (top panels) and imaginary (lower panels) parts of the $l=0$ (monopole) mode of mini-BSs as a function of the central value of the scalar field (left panels) and of the star compactness (right panels). The vertical dashed line indicates the maximum mass configuration. The mode becomes unstable ($\sigma_I>0$) when the density exceeds the critical value corresponding to the maximum mass (see Fig.~\ref{fig:zoom_l0}).}%
\label{fig:modes_phic_l0}%
\end{figure}
We can now apply the above procedure to compute the QNMs of a BS. A comparison between results obtained with the direct integration procedure, as in Ref.~\cite{Macedo:2013jja}, and with the continued-fraction method for some BS configurations is shown in Table~\ref{tab:modes}. Because of to the small imaginary part of the modes, the divergence of the QNMs is weak and the agreement between the two methods is remarkable. In fact, this good agreement between the two methods is expected because the direct integration method is suitable for computing the least damped modes.

\begin{figure}
\hspace*{-1cm}
\includegraphics[width=0.65\textwidth]{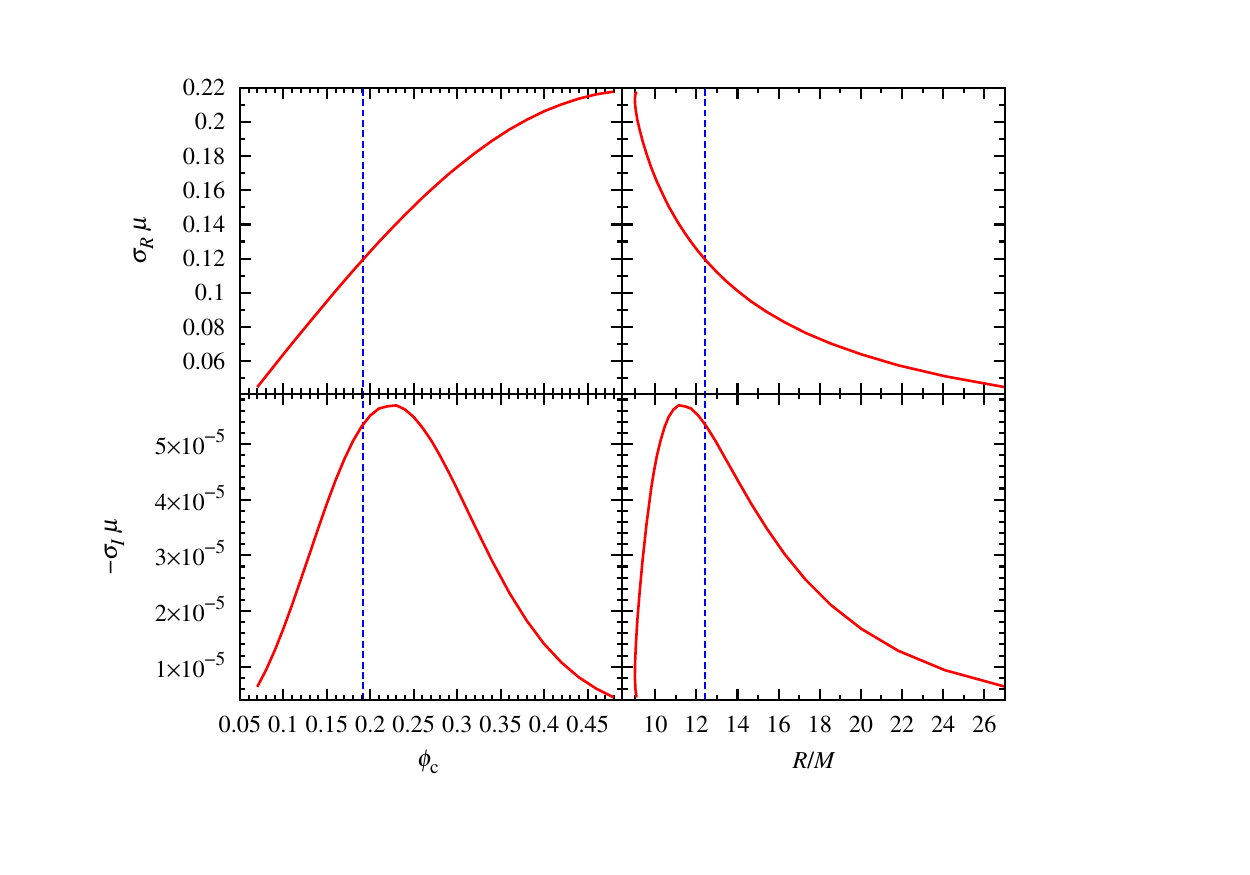}%
\vspace*{-1cm}
\caption{Same as Fig. \ref{fig:modes_phic_l0}, but for $l=2$ (quadrupole).}%
\label{fig:modes_phic_l2}%
\end{figure}
In the case of BSs, the advantage of the continued-fraction method over the direct integration is in the construction of the background solutions. For the direct integration method to work properly, we need to integrate from a point in which ${\bf V}(r)$ has approximately a constant value (or zero). This integration from numerical infinity can be strongly contaminated by numerical errors, such that one generally has to compute the sum in Eq.~\eqref{eq:expansioninf} up to, typically, $N_0\sim15$, and the background solutions need to be constructed up to a point far away from the star. On the other hand, the only requirement for the continued-fraction method to work properly is that the background scalar field is small enough at $R_{\rm ext}$, and therefore that the continued-fraction solutions can be constructed in a point relatively close to the star.

\begin{figure}%
\includegraphics[width=0.9\columnwidth]{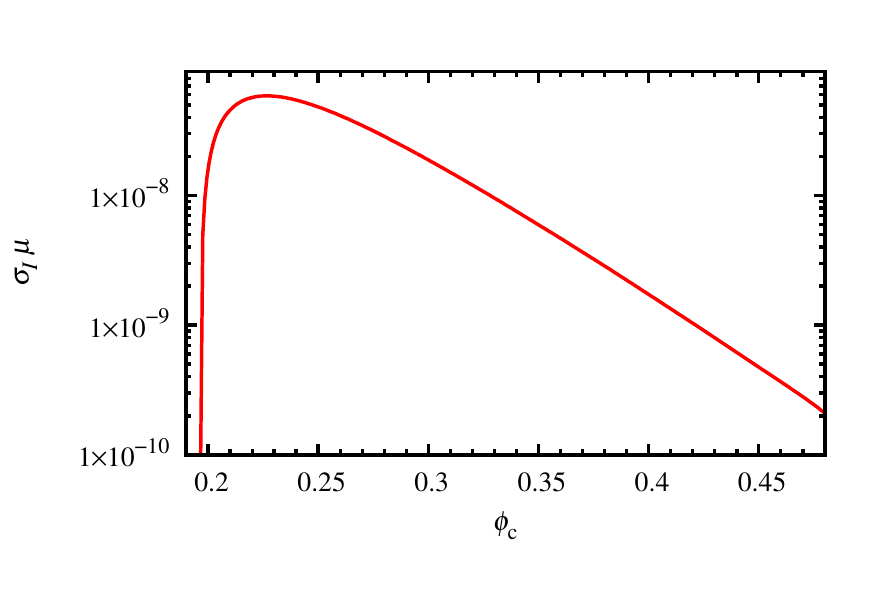}%
\vspace*{-1cm}
\caption{A zoom in the range of the unstable $l=0$ modes of mini-BSs.}%
\label{fig:zoom_l0}%
\end{figure}

By using the continued-fraction method, we can now compute the modes of mini-BSs as a function of the central density and radius of the star, defined as the radial point $r$ which contains $99\%$ of the star's total mass \cite{Macedo:2013jja}. The results are shown in Figs.~\ref{fig:modes_phic_l0} and \ref{fig:modes_phic_l2}. The monopole mode ($l=0$, Fig. \ref{fig:modes_phic_l0}) exists due to the coupling between the scalar and gravitational perturbations \cite{Yoshida:1994xi,Macedo:2013jja}. In Fig. \ref{fig:modes_phic_l2}, we observe that the quadrupole mode is stable in the considered range of the central density. On the other hand, the monopole mode becomes unstable ($\sigma_I>0$; cf. Fig.~\ref{fig:zoom_l0}) for central densities $\phi\gtrsim \phi_c^{\rm crit}$, where $\phi_c^{\rm crit}\sim 0.1916$ is the central density corresponding to the maximum mass configuration~\cite{Schunck:2003kk}. This is in accordance with previous works which considered the radial stability of BSs~\cite{Lee:1988av,Kusmartsev:1990cr,Hawley:2000dt}.

\section{Discussion and final remarks} \label{sec:discusscf}

We extended two different methods for computing the QNMs of compact objects in the presence of interacting fields. One of the methods is based on a direct integration of the differential equations and the other is based on an analytical continued-fraction scheme for describing the perturbations. The methods can be applied in many astrophysical situations, e.g., to compute the QNMs of relativistic stars surrounded by extra (electromagnetic or exotic) fields.

We applied the direct integration and continued-fraction methods to compute the polar QNMs of BSs. For the modes analyzed, the continued-fraction method is in excellent agreement with the direct integration one. Moreover, we computed the modes as a function of the central value of the scalar field, showing that the monopole mode becomes unstable for configurations beyond the maximum mass.

The techniques presented in this paper can be useful for the computation of QNMs in complicated configurations. One limitation of the continued-fraction method, however, is manifest when dealing with lesser compact interacting fields, i.e., configurations with the longer length scale $l_i\gg R$. In that case, the divergence of the fields usually grows rapidly as $r$ increases, and numerical errors can contaminate the accuracy of the solution. One possible way to circumvent that problem is to mix the direct integration method with the continued-fraction one---for instance, one can use the continued-fraction expansion as an outer boundary condition for the integration from infinity, thus decreasing the errors introduced when using the expansion~\eqref{eq:expansioninf}.

\begin{acknowledgments}

C.M. and L.C. would like to thank Conselho Nacional de Desenvolvimento Cient\'ifico e Tecnol\'ogico (CNPq), Coordena\c{c}\~ao de Aperfei\c{c}oamento de Pessoal de N\'ivel Superior (CAPES), and Funda\c{c}\~ao Amaz\^onia de Amparo a Estudos e Pesquisas do Par\'a (FAPESPA).
P.P. was supported by the European Community through
the Intra-European Marie Curie Contract No.~AstroGRAphy-2013-623439 and by FCT-Portugal through the Project No. IF/00293/2013.
V.C. acknowledges financial support provided under the European
Union's FP7 ERC Starting Grant ``The dynamics of black holes: testing
the limits of Einstein's theory'' Grant Agreement No. DyBHo--256667,
and H2020 ERC Consolidator Grant ``Matter and strong-field gravity: New frontiers in Einstein's theory'' Grant Agreement No. MaGRaTh--64659.
C.M. thanks the Perimeter Institute for the kind hospitality.
Research at the Perimeter Institute is supported by the Government of Canada 
through Industry Canada and by the Province of Ontario through the Ministry
of Economic Development and Innovation.
This work was supported by the NRHEP 295189 FP7-PEOPLE-2011-IRSES Grant, and by FCT-Portugal through Projects No.
PTDC/FIS/098025/2008, No. PTDC/FIS/098032/2008, No. PTDC/FIS/116625/2010, and No.
CERN/FP/116341/2010.
This work was also partly funded through H2020 ERC Consolidator Grant ``Matter and strong-
field gravity: New frontiers in Einstein’s theory'' Grant Agreement No. MaGRaTh–64659.
%
\end{acknowledgments}

\bibliography{references}
\end{document}